\documentclass[twoside,reqno]{qts-proc}
\usepackage{epsfig,cite}
\usepackage{amssymb,amsmath}
\usepackage{pstricks,bbm}
\usepackage{times}
\DeclareMathAlphabet{\boldmathe}{T1}{cmr}{bx}{it}
\begin{document}
\sloppy \raggedbottom
\setcounter{page}{1}
\newcommand{\matrixx}[1]{\begin{pmatrix}#1\end{pmatrix}}
\newcommand{\mtxt}[1]{\quad\text{#1}\quad}
\def\pa{\partial}
\def\psida{\psi^\dagger}
\def\id{\mathbbm{1}}
\def\bull{$\bullet$}
\def\be{\begin{equation}}
\def\ee{\end{equation}}
\def\HB{H_\text{B}}
\def\HF{H_\text{F}}
\def\vcC{\boldmathe{C}}
\def\vcJ{\boldmathe{J}}
\def\vcK{\boldmathe{K}}
\def\vcL{\boldmathe{L}}
\def\vcW{\boldmathe{W}}
\def\vcS{\boldmathe{S}}
\def\vcp{\boldmathe{p}}
\def\vcr{\boldmathe{r}}
\def\vcx{\boldmathe{x}}
\def\cC{{\mathcal C}}
\def\cH{{\cal H}}
\def\cN{{\cal N}}
\def\cQ{{\cal Q}}
\def\cQd{{\cal Q}^\dagger}
\def\cS{{\cal S}}
\def\pa{\partial}
\def\psid{\psi^\dagger}
\def\psia{\psi_a}
\def\psiad{\psi_a^\dagger}
\def\psib{\psi_b}
\def\psibd{\psi_b^\dagger}
\def\vac{\vert 0\rangle}
\def\setR{\mathbbm{R}}
\def\setC{\mathbbm{C}}
\def\setN{\mathbbm{N}}
\newcommand{\mbfgr}[1]{\textit{\mbox{\boldmath$#1$}}}
\def\vcpsi{\mbfgr{\psi}}
\newpage
\setcounter{figure}{0}
\setcounter{equation}{0}
\setcounter{footnote}{0}
\setcounter{table}{0}
\setcounter{section}{0}
\title{Algebraic Solution of the Supersymmetric Hydrogen Atom}

\runningheads{Wipf}{Supersymmetric H-Atom}

\begin{start}

\author{Andreas Wipf}{1},
\coauthor{Andreas Kirchberg}{2},
\coauthor{Dominique L\"ange}{3}

\address{FS-Univ. Jena, Max-Wien-Platz 1, 
Theor. Phys. Institut, D-07743 Jena}{1}
\address{Reichenberger Stra{\ss}e 9, 01129 Dresden}{2}
\address{LM-Univ. M\"unchen,
Sektion Physik, Theresienstr. 37, D-80333 M\"unchen}{3}

\begin{Abstract}
The $\cN\!=\!2$ supersymmetric extension of the 
\textsc{Schr\"odinger-Hamilton}ian with $1/r$-potential in $d$ dimension is
constructed. The system admits a supersymmetrized \textsc{Laplace-Runge-Lenz} 
vector which extends the rotational $SO(d)$ symmetry to a 
hidden $SO(d+1)$ symmetry. It is used to determine 
the discrete eigenvalues with their degeneracies and 
the corresponding bound state wave functions. 
\end{Abstract}
\end{start}


\section{Classical motion in \emph{Newton/Coulomb} potential}
For a closed system of two non-relativistic point
masses interacting via a central force the angular
momentum $\vcL$ of the relative motion is conserved
and the motion is always in the plane perpendicular to
$\vcL$. If the force is derived from a 
$1/r$-potential, there is an additional conserved quantity: the 
\textsc{Laplace-Runge-Lenz}\footnote{A more suitable name for
this constant of motion would be 
\textsc{Hermann-Bernoulli-Laplace}
vector, see \cite{goldstein}.} vector,
\[
\vcC=\frac{1}{m}\,\vcp\times\vcL-\frac{e^2}{r}\,\vcr.
\]
This vector is perpendicular to $\vcL$ and
points in the direction of the semi-major axis.
For the hydrogen atom the corresponding Hermitian vector
operator has the form
\be
\vcC=\frac{1}{2m}\left(\vcp\times\vcL-\vcL\times\vcp\right)
-\frac{e^2}{r}\vcr \label{rungelenz}
\ee
with reduced mass $m$ of the proton-electron
system.
By exploiting the existence of this conserved vector operator,
\textsc{Pauli} calculated the spectrum of the hydrogen atom by purely
algebraic means \cite{pauli,fock}. He noticed that 
the angular momentum $\vcL$ together with the vector operator
\be
\vcK=\sqrt{\frac{-m}{2H}}\,\vcC,
\ee
which is well-defined and Hermitian on bound states with
negative energies,
generate a hidden $SO(4)$ symmetry algebra,
\begin{eqnarray}
[L_a,L_b]&=&i\epsilon_{abc}L_c\,,\cr
[L_a,K_b]&=&i\epsilon_{abc}K_c\,,\cr
[K_a,K_b]&=&i\epsilon_{abc}L_c\,,\label{algebra}
\end{eqnarray}
and that the \textsc{Hamilton}-Operator
can be expressed in terms of $\cC_{(2)}=\vcK^2+\vcL^2$, one of the
two second-order \textsc{Casimir} operators of this algebra,
as follows
\be
H=-\frac{me^4}{2}\frac{1}{\cC_{(2)}+\hbar^2}.\label{ham3d}
\ee
One also notices that the second \textsc{Casimir}
operator  $\tilde \cC_{(2)}=\vcL\cdot\vcK$ vanishes and arrives
at the bound state energies by purely group
theoretical methods. The existence
of the conserved vector $\vcK$ also explains the 
accidental degeneracy of the hydrogen spectrum.

We generalize the \textsc{Coulomb}-problem to $d$
dimensions by keeping the $1/r$-potential.
Distances are measured in units of the
reduced \textsc{Compton} wavelength,
such that the \textsc{Schr\"odinger}-operator takes the form
\be
H=p^2-\frac{\eta}{r}\;,\quad  p_a
=\frac{1}{i}\, \pa_a\;, \quad a=1,\dots,d\;.\label{hamilton}
\ee
$\eta$ is twice the fine structure constant.
Energies are measured in units of $mc^2/2$.

The Hermitian generators $L_{ab}=x_a p_b -x_b p_a$ of the
rotation group satisfy the 
familiar $so(d)$ commutation relations
\be
[ L_{ab}, L_{cd} ] = i \left(\delta_{ac} L_{bd} 
+ \delta_{bd} L_{ac} - \delta_{ad} L_{bc} - \delta_{bc}
L_{ad} \right) \;.\label{sonalgebra}
\ee
It is not very difficult to guess the
generalization of the \textsc{Laplace-Runge-Lenz} vector (\ref{rungelenz})
in $d$ dimensions \cite{sudarshan},
\be
C_a = L_{ab} p_b + p_b L_{ab} -\frac{\eta x_a}{r} \;.
\ee
These operators commute with $H$ in (\ref{hamilton}) 
and form a $SO(d)$-vector,
\be
[L_{ab},C_c]=i (\delta_{ac}C_b-\delta_{bc}C_a)\;.
\ee
The commutator of $C_a$ and $C_b$
is proportional to the angular momentum,
\be
[ C_a, C_b ] = -4i L_{ab} H\;.
\ee
Now one proceeds as in three dimensions and defines
on the negative energy subspace of $L_2(\setR^d)$ 
the Hermitian operators
\be
K_a =\frac{1}{2} \frac{C_a}{\sqrt{-H}} \mtxt{ with }[K_a,K_b] 
= i L_{ab} \;.
\ee
The operators $\{L_{ab},K_a\}$
form a closed symmetry algebra and can be combined 
to form generators $L_{AB}$ of the orthogonal 
group\footnote{For scattering states $(E>0)$ a 
similar redefinition leads to generators 
of the Lorentz group $SO(d,1)$.
Here we are interested in bound states and will not 
further discuss this possibility.} $SO(d+1)$,
\be
L_{AB} = \left(\begin{array}{c|c} L_{ab} & K_a \\ \hline 
-K_b & 0 \end{array}\right).
\ee
They obey the commutation relations (\ref{sonalgebra}) 
with indices running from $1$ to $d+1$. 

One finds a relation similar to (\ref{ham3d}) by solving
\[
 C_a C_a =-4K_aK_a H
=\eta^2 + \left(2 L_{ab} L_{ab} +(d-1)^2\right) H
\]
for the Hamiltonian,
\be
H =p^2-\frac{\eta}{r}=
 - \frac{\eta^2}{(d-1)^2+4 \cC_{(2)}}.\label{bos12}
\ee
$\cC_{(2)}$ is the second-order \textsc{Casimir} operator of the 
dynamical symmetry group,
\be
\mathcal{C}_{(2)}= \frac 12 L_{AB} L_{AB} 
=\frac 12 L_{ab} L_{ab} + K_aK_a \;.
\ee
It remains to find the admitted
irreducible representations of $SO(d+1)$.
In three dimensions they are
fixed by the condition $\tilde \cC_{(2)}=0$
on the \textsc{Casimir} operator not entering
the relation (\ref{ham3d}).
In $d=2n\!-\!1$ and $d=2n$ dimensions there are 
$n$ \textsc{Casimir} operators of the dynamical symmetry
group and we expect $n-1$ conditions.
The analysis in 
\cite{wipfkirch2} lead to the following results:
\begin{itemize}
\itemsep=0pt
\item Only the completely symmetric representations of $SO(d+1)$ 
are realized.
\item As in three dimensions the energies, degeneracies and eigenfunctions are 
determined by group-theoretic methods.
\end{itemize}
\section{Susy Quantum Mechanics}
The \textsc{Hilbert}-Space of a supersymmetric 
system is the sum of its bosonic and fermionic subspaces, 
$\cH=\cH_{\text{B}}\oplus \cH_{\text{F}}$.
Let $A$ be a linear operator $\cH_{\text{F}}\to \cH_{\text{B}}$.
We shall use a block notation such that the vectors
in $\cH_{\text{B}}$ have upper and those in
$\cH_{\text{F}}$ lower components,
\[
\vert\psi\rangle=\matrixx{\vert\psi_{\text{B}}\rangle\cr \vert\psi_{\text{F}}\rangle}.
\]
Then the nilpotent 
\emph{supercharge} and its adjoint take the forms
\be
\cQ=\matrixx{0&A\cr 0&0},\quad
\cQd=\matrixx{0&0\cr A^\dagger&0}\Longrightarrow
\{\cQ,\cQ\}=0.
\ee
The block-diagonal super-\textsc{Hamilton}ian
\be
H\equiv\{\cQ,\cQd\}=\matrixx{AA^\dagger & 0\cr 0& A^\dagger A }=
\matrixx{\HB&0\cr 0&\HF},
\ee
commutes with the supercharge and the (fermion) number operator
\[
\text{N}=\matrixx{0&0\cr 0&1}.
\]
Bosonic states have $\text{N}\!=\!0$ 
and fermionic states $\text{N}\!=\!1$.
The supercharge and its adjoint decrease and increase 
this conserved number by one.

In most applications in quantum mechanics
$A$ is a first order differential operator
\be
A=i\pa_x+iW(x)
\ee
and yields the isospectral partner-\textsc{Hamilton}ians 
\be
\HB=p^2+V_{\text{B}}\;,\quad \HF=p^2+V_{\text{F}},
\quad\text{with}\quad
V_{\text{B}/\text{F}}=W^+\pm W'.
\ee
Such one-dimensional systems were introduced by
\textsc{Nicolai} and \textsc{Witten} some
time ago \cite{nicolai76,witten81}.
See the texts \cite{cooper,wipfgenf}
for a discussion of such models 
and in particular their relation to isospectral 
deformations and integrable systems.
\section{SQM in Higher Dimensions}
Supersymmetric quantum mechanical systems 
also exist in higher dimensions \cite{witten81,andrianov}. 
The construction is motivated by the following 
rewriting of the supercharge 
\[
\cQ=\psi\otimes A\mtxt{and}\cQd=\psi^\dagger\otimes A^\dagger
\]
containing the \emph{fermionic} operators
\[
\psi=\matrixx{0&1\cr 0&0}\mtxt{and}
\psi^\dagger=\matrixx{0&0\cr 1&0}
\]
with anti-commutation relations
\[
\{\psi,\psi\}=\{\psi^\dagger,\psi^\dagger\}=0\mtxt{and}
\{\psi,\psi^\dagger\}=\id.
\]
In \cite{andrianov} this construction has been
generalized to $d$ dimensions. Then one
has $d$ fermionic annihilation operators $\psi_k$ 
and $d$ creation operators $\psi^\dagger_k$,
\be
\{\psi_k,\psi_\ell\}=\{\psida_k,\psida_\ell\}=0\mtxt{and}
\{\psi_k,\psida_\ell\}=\delta_{k\ell},\quad k,\ell=1,\dots,d.
\ee
For the supercharge one makes the ansatz
\[
\cQ=i\sum \psi_k \left(\pa_k+W_k(\vcx)\right).
\]
It is \emph{nilpotent} (i.e. $\cQ^2=0$)
if and only if $\pa_kW_\ell-\pa_\ell W_k=0$ 
holds true. Locally this integrability condition is 
equivalent to the existence of a potential
$\chi(x)$ with $W_k=\pa_k\chi$. Thus we are lead to
the following \emph{nilpotent} supercharge
\be
\cQ=
e^{-\chi}\cQ_0\,e^\chi\;\mtxt{with}
\cQ_0=i\sum \psi_k\pa_k\,.\label{supercharge}
\ee
It acts on elements of the \textsc{Hilbert}-space 
\[
\cH=L_2(\setR^d)\otimes \setC^{2^d},
\]
which is graded by the 'fermion-number' operator $\text{N}=\sum \psiad\psia\,$,
\be
\cH=\cH_0\oplus \cH_1\oplus \ldots \oplus \cH_d,\qquad
\text{N}\big\vert_{\cH_p}=p\id.
\ee
A state in $\cH_p$ has the expansion
\be
\Psi=\sum f_{a_1\dots a_p}(x)\;\vert a_1\dots a_p\rangle,\quad
\vert a_1\dots a_p\rangle=\psid_{a_1}\cdot\cdot\cdot\psid_{a_p}\vac
\ee
with antisymmetric $f_{a_1\dots a_p}$.
$\cQ$ decreases $\text{N}$ by one
and its adjoint increases it by one.
It follows that the super-\textsc{Hamilton}ian 
\begin{eqnarray}
H=\{\cQ,\cQd\}&=&H_0\otimes \id_{2^d}
-2\sum\psida_k \psi_\ell\,\pa_k\pa_\ell\chi\cr
&=& H_d\otimes \id_{2^d}
+2\sum\psi_k \psida_\ell\,\pa_k\pa_\ell\chi\label{susyham}
\end{eqnarray}
preserves the 'fermion-number'. The operators
in the extreme sectors,
\begin{eqnarray}
H_0\equiv H\big\vert_{\cH_0}&=&-\triangle+(\nabla\chi,\nabla\chi)+\triangle \chi\cr
H_d\equiv H\big\vert_{\cH_d}&=&-\triangle+(\nabla\chi,\nabla\chi)-\triangle \chi.
\end{eqnarray}
are ordinary \textsc{Schr\"odinger}-operators, whereas
the restriction of $H$ to any other sector is
a matrix-\textsc{Schr\"odinger}-operator, 
\[
H_p\equiv H\big\vert_{\cH_p}:\quad 2^{\binom{d}{p}}\times 2^{\binom{d}{p}}
\;-\;
\text{matrix.}
\]
Due to the nilpotency of $Q$ and $[Q,H]=0$ one has
a \textsc{Hodge}-type decomposition of the
\textsc{Hilbert}-space \cite{wipfkirch2},
\be
\cH=\cQ\cH \oplus \cQd\cH \oplus \hbox{Ker}\,H\label{hodge}\,.
\ee
Actually, the graded \textsc{Hilbert}-space is a $\cQ$-complex
of the following structure,
\begin{center}
\psset{unit=1mm,linewidth=.6pt}
\begin{pspicture}(0,-4)(100,15)
\rput(5,5){$\cH_0$}
\psline{->}(9,6)(21,6)
\psline{<-}(9,4)(21,4)
\rput(15,9){$\cQd$}
\rput(15,1){$\cQ$}
\rput(25,5){$\cH_1$}
\psline{->}(29,6)(41,6)
\psline{<-}(29,4)(41,4)
\rput(35,9){$\cQd$}
\rput(35,1){$\cQ$}
\rput(45,5){$\cH_2$}
\psline{->}(49,6)(61,6)
\psline{<-}(49,4)(61,4)
\rput(55,9){$\cQd$}
\rput(55,1){$\cQ$}
\multirput(64,5)(4,0){4}{\psdot}
\psline{->}(79,6)(89.5,6)
\psline{<-}(79,4)(89.5,4)
\rput(85,9){$\cQd$}
\rput(85,1){$\cQ$}
\rput(95,5){$\cH_d$}
\end{pspicture}
\end{center}
Similarly as in the one-dimensional case one
has a pairing of all $H$-eigenstates with
non-zero energy. Every excited state
is degenerate and the eigenfunctions are
mapped into each other by $\cQ$ and its
adjoint. The situation is depicted in
the following figure,
\begin{center}
\psset{xunit=.9mm,yunit=1mm}
\begin{pspicture}(-10,-10)(100,80)
\small
\rput(-5,57){$E$}
\psline{->}(0,10)(0,62)
\psline(30,10)(30,62)
\psline(60,10)(60,62)
\psline(90,10)(90,62)
\psline(0,10)(100,10)
\rput(0,70){$\cH_0$}
\psline{->}(5,71)(25,71)
\rput(15,75){$\cQd$}
\psline{->}(25,69)(5,69)
\rput(15,65){$\cQ$}
\rput(30,70){$\cH_1$}
\psline{->}(35,71)(55,71)
\rput(45,75){$\cQd$}
\psline{->}(55,69)(35,69)
\rput(45,65){$\cQ$}
\rput(60,70){$\cH_2$}
\psline{->}(65,71)(85,71)
\rput(75,75){$\cQd$}
\psline{->}(85,69)(65,69)
\rput(75,65){$\cQ$}
\rput(90,70){$\cH_3$}
\rput[l](95,70){$\dots$}
\psline(0,20)(10,20)\psline(20,20)(30,20)\psline[linestyle=dotted](10,20)(20,20)
\psline(0,35)(10,35)\psline(20,35)(30,35)\psline[linestyle=dotted](10,35)(20,35)
\psline(0,42)(10,42)\psline(20,42)(30,42)\psline[linestyle=dotted](10,42)(20,42)
\psline(0,50)(10,50)\psline(20,50)(30,50)\psline[linestyle=dotted](10,50)(20,50)
\psline(0,56)(10,56)\psline(20,56)(30,56)\psline[linestyle=dotted](10,56)(20,56)
\psline(30,25)(40,25)\psline(50,25)(60,25)\psline[linestyle=dotted](40,25)(50,25)
\psline(30,39)(40,39)\psline(50,39)(60,39)\psline[linestyle=dotted](40,39)(50,39)
\psline(30,46)(40,46)\psline(50,46)(60,46)\psline[linestyle=dotted](40,46)(50,46)
\psline(30,53)(40,53)\psline(50,53)(60,53)\psline[linestyle=dotted](40,53)(50,53)
\psline(30,58)(40,58)\psline(50,58)(60,58)\psline[linestyle=dotted](40,58)(50,58)
\psline(60,27)(70,27)\psline(80,27)(90,27)\psline[linestyle=dotted](70,27)(80,27)
\psline(60,41)(70,41)\psline(80,41)(90,41)\psline[linestyle=dotted](70,41)(80,41)
\psline(60,48)(70,48)\psline(80,48)(90,48)\psline[linestyle=dotted](70,48)(80,48)
\psline(60,54)(70,54)\psline(80,54)(90,54)\psline[linestyle=dotted](70,54)(80,54)
\psline(60,59)(70,59)\psline(80,59)(90,59)\psline[linestyle=dotted](70,59)(80,59)
\psline[linestyle=dotted](15,10)(15,60)
\psline[linestyle=dotted](45,10)(45,60)
\psline[linestyle=dotted](75,10)(75,60)
\multirput(0,0)(30,0){3}{\psframe(0,0)(15,10)\rput(7.5,5){$\cQ\cH$}}
\psset{fillstyle=solid,fillcolor=lightgray}
\multirput(15,0)(30,0){3}{\psframe(0,0)(15,10)\rput(7.5,5){$\cQd\cH$}}
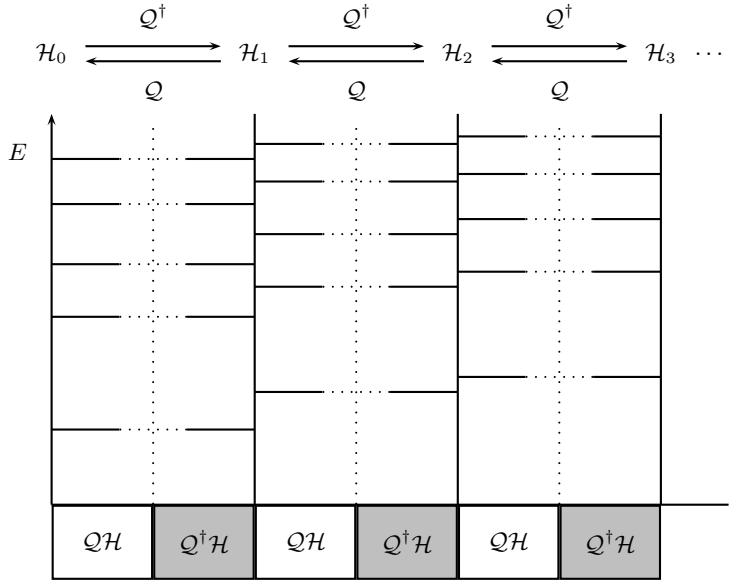
\rput(50,-8){pairing of states with $E>0$}
\end{pspicture}\\
\end{center}
\section{The supersymmetric H-Atom}
We supersymmetrized the
H-atom along these lines and showed that 
it admits supersymmetric generalizations 
of the angular momentum and \textsc{Laplace-Runge-Lenz} 
vector \cite{wipfkirch2}. 
As for the ordinary \textsc{Coulomb} problem the 
hidden $SO(d+1)$-symmetry allows for a purely algebraic 
solution. Here we discuss the construction for the
$3$-dimensional system and sketch 
the generalization to arbitrary dimensions.

To construct the supersymmetrized
H-atom in $3$ dimensions we choose $\chi=-\lambda r$ in (\ref{supercharge}) and
obtain the super-\textsc{Hamilton}ian
\cite{wipfkirch2}
\be
H=(-\triangle+\lambda^2)\id_8-\frac{2\lambda}{r}B,\quad
B=\id-\text{N}+S^\dagger S,\quad S=\hat\vcx\cdot \vcpsi\label{susyH1}
\ee
on the \textsc{Hilbert}-space
\be
\cH=L_2(\setR^3)\times \setC^8=\cH_0\oplus \cH_1\oplus\cH_2\oplus\cH_3.
\ee
We defined the triplet $\vcpsi$ containing the $3$
annihilation operators $\psi_1,\psi_2,\psi_3$.
States in $\cH_0$ are annihilated by $S$
and states in $\cH_3$ by $S^\dagger$. With
$\{S^\dagger,S\}=\id$ we find the following
\textsc{Hamilton}-operators in these extreme subspaces,
\begin{eqnarray*}
H_0=-\triangle+\lambda^2-\frac{2\lambda}{r},\cr
H_3=-\triangle+\lambda^2+\frac{2\lambda}{r}.
\end{eqnarray*}
$H_0$ describes the proton-electron and $H_3$
the proton-positron system.

The conserved angular momentum contains a spin-type term,
\be
\vcJ=\vcL+\vcS=\vcx\wedge\vcp-i\vcpsi^\dagger\wedge\vcpsi.\label{susyH5}
\ee
The operators $\vcx$ and $\vcpsi$ are both vectors
such that $S$ and $B$ in  (\ref{susyH1}) commute
with this total angular momentum.
To find the susy extension of the \textsc{Runge-Lenz} 
vector is less simple. It reads
\cite{wipfkirch2}
\be
\vcC=\vcp\wedge\vcJ-\vcJ\wedge\vcp-2\lambda\, \hat\vcx B
\ee
with $\vcJ$ from (\ref{susyH5}) and $B$ from (\ref{susyH1}). 
The properly normalized vector
\be
\vcK=\frac{1}{2}\frac{\vcC}{\sqrt{\lambda^2-H}}
\ee
together with $\vcJ$ form an $SO(4)$ symmetry algebra
on the subspace of bound states for
which $H<\lambda^2$.

To solve for the spectrum we would like to find a 
relation similar to (\ref{ham3d}). However, one soon realizes that
there is no algebraic relation between the
conserved operators $\id, \text{N},\vcJ^2,
\vcK^2$ and $H$. However, we can prove
the equation
\begin{eqnarray}
\lambda^2\cC_{(2)}=\vcK^2 H
&+&\left(\vcJ^2 + (1-\text{N})^2 \right) \cQ\cQd\cr
&+&\left(\vcJ^2+(2-\text{N})^2 \right) \cQd \cQ \;,\label{susyH9}
\end{eqnarray}
where $\cC_{(2)}$ is the second-order \textsc{Casimir}
(\ref{ham3d}). This relation is sufficient to obtain the
energies
since each of the three subspaces in the 
\textsc{Hodge}-decomposition
(\ref{hodge}) is left invariant by $H$
and thus we may diagonalize it on each subspace separately.
Since $H\vert_{\cQ\cH}=\cQ\cQd$ 
and $H\vert_{\cQd\cH}=\cQd \cQ$ we can solve 
(\ref{susyH9}) for $H$ in both subspaces,
\begin{eqnarray}
H\big\vert_{\cQ\cH}=\lambda^2\,\frac{\cC_{(2)}}{(1-\text{N})^2+\cC_{(2)}}\,,\cr
H\big\vert_{\cQd\cH}=\lambda^2\,\frac{\cC_{(2)}}{(2-\text{N})^2+\cC_{(2)}}\,.
\end{eqnarray}
States with zero energy are annihilated by both
$\cQ$ and $\cQd$, and according to (\ref{susyH9}) 
the second-order \textsc{Casimir} must vanish on these modes,
such that
\[
\cC_{(2)}\big\vert_{\hbox{Ker}\,H}=0\;.\]
We conclude that every supersymmetric ground state of $H$
is an $SO(4)$ singlet.

In the figure below we have plotted
the spectrum of the supersymmetric
$H$-atom in $3$ dimensions. The bound states reside in the sectors
with fermion numbers $0$ and $1$. 
In the sectors with fermion numbers $2$ and $3$
there are only scattering states.
All bound states transform according to
the symmetric representations of $SO(4)$. 
This is particular to $3$ dimensions.
The energies with degeneracies
and the wave functions for all
bound states can be found in \cite{wipfkirch2}.
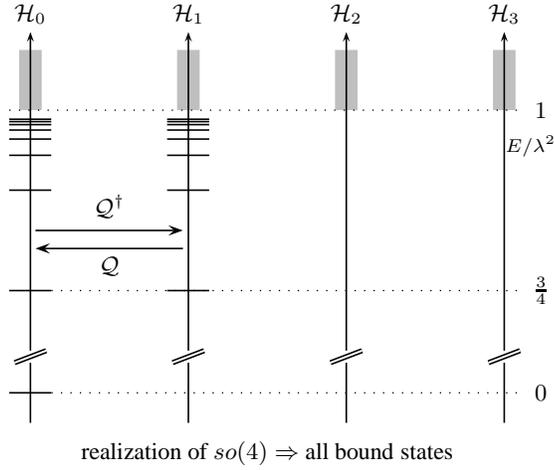
\begin{center}
\psset{xunit=.7cm,yunit=.8cm,linewidth=0.6pt,dotsize=1mm}
\begin{pspicture}(0,0)(11,8.5)
\multirput(1,1)(3,0){4}{\psline(0,0)(0,1)}
\psline(.6,1.5)(1.4,1.5)
\psline[linestyle=dotted](1.4,1.5)(10.4,1.5)
\multirput(1,2.1)(3,0){4}{\psline[doubleline=true](-.3,-.1)(.3,.1)}
\multirput(1,6.2)(3,0){4}{
\psline[linewidth=.3,linecolor=lightgray](0,0)(0,1)}
\multirput(1,2.2)(3,0){4}{\psline{->}(0,0)(0,5.3)}
\psline[linestyle=dotted](1.4,3.2)(10.4,3.2)
\psline[linestyle=dotted](.6,6.2)(10.4,6.2)
\multirput(0,0)(3,0){2}{
\psline(0.6,3.2)(1.4,3.2)
\psline(0.6,4.87)(1.4,4.87)
\psline(0.6,5.45)(1.4,5.45)
\psline(0.6,5.72)(1.4,5.72)
\psline(0.6,5.87)(1.4,5.87)
\psline(0.6,5.96)(1.4,5.96)
\psline(0.6,6.01)(1.4,6.01)
\psline(0.6,6.05)(1.4,6.05)}
\small
\rput(10.7,1.5){$0$}
\rput(10.7,3.2){$\frac{3}{4}$}
\rput(10.7,6.2){$1$}
\rput(10.5,5.6){\scriptsize$E/\lambda^2$}
\rput(1,7.8){$\cH_0$}
\rput(4,7.8){$\cH_1$}
\rput(7,7.8){$\cH_2$}
\rput(10,7.8){$\cH_3$}
\psline[arrowsize=1.4mm]{->}(1.1,4.2)(3.9,4.2)
\rput(2.5,4.6){$\cQd$}
\psline[arrowsize=1.4mm]{->}(3.9,3.9)(1.1,3.9)
\rput(2.5,3.6){$\cQ$}
\rput(5.5,.5)
{realization of $so(4)\Rightarrow$ all bound states}
\end{pspicture}
\end{center}

\section{Higher dimensions}
The super-\textsc{Hamilton}ian (\ref{susyham}) 
with $\chi\!=-\lambda r$ describes a supersymmetrized
\textsc{Coulomb}-problem in $d$ dimensions. 
As in $3$ dimensions it can be solved with the 
help of a supersymmetrized angular momentum
and \textsc{Runge-Lenz} vector generating
a dynamical symmetry $SO(d+1)$.
The supersymmetric extension of the angular momenta
reads
\be
J_{ab}=L_{ab}+S_{ab}\mtxt{with}
S_{ab}=\frac{1}{i}\left(\psid_a\psi_b-\psid_b\psi_a\right).
\ee
The supercharge, \textsc{Hamilton}ian and $\cS=\hat{\vcx}\cdot\vcpsi$ 
are scalars with respect to the rotations generated
by the $J_{ab}$. The supersymmetric extension of 
\textsc{Laplace-Runge-Lenz} vector
\be
C_a = J_{ab}p_b+p_b J_{ab}-2\lambda \hat{x}_a B
\ee
and the super-\textsc{Hamilton}ian
\be
H= - \triangle + \lambda^2 - \frac {2\lambda }{r}\,B
\ee
both contain the scalar operator
\be
B =\frac{1}{2}(d-1)\id -\text{N}+\cS^\dagger \cS.
\ee
Again the \textsc{Fock-Bargmann} symmetry group $SO(d+1)$ 
is generated by
\[
L_{AB} = \matrixx{L_{ab} & K_a\cr -K_b & 0},\quad
K_a = \frac {C_a}{\sqrt{4(\lambda^2 - H)}},
\]
and the second-order \textsc{Casimir}
\be
\cC_{(2)} = \frac{1}{2}J_{AB} J_{AB},
\ee
together with $\lambda,d,\text{N}$ enter the formulas for
\[
H\big\vert_{\cQ\cH}\mtxt{and}H\big\vert_{\cQd\cH}.
\]
The analysis parallels the one in $3$ dimensions.
To find the allowed representations one uses
the branching-rules from the dynamical symmetry $SO(d+1)$
to the rotational symmetry $SO(d)$ generated by the $J_{ab}$.
Only those representation for which the \textsc{Young}-diagram
has exactly one row and exactly one column give rise to normalizable
states. The construction of the bound state wave function
uses the realization of the \textsc{Cartan}- and step 
operators $H_\alpha,E_\alpha$ as differential operators. This way one
finds the highest weight state in each representation \cite{wipfkirch2}.

\section{Conclusions}
We have succeeded in supersymmetrizing the celebrated 
construction of \textsc{Pauli, Fock} and \textsc{Bargmann}.
For the \textsc{Coulomb}-problem with extended
$\cN=2$ supersymmetry we have found
the conserved angular momentum and conserved
\textsc{Runge-Lenz} vector. Together they generate
the \textsc{Fock-Bargmann} symmetry group $SO(d+1)$. 
A general relation of the type
\be
\cQ\cQd=f_1\left(\lambda,d,\textbf{N},\cC_{(2)}\right)\mtxt{and}
\cQd\cQ=f_2\left(\lambda,d,\textbf{N},\cC_{(2)}\right)
\ee
has been derived which allows for an algebraic
treatment of the supersymmetrized hydrogen atom
in $d$ dimensions.
The energies depend on the fine structure
constant, the dimension of space, the fermion
number and the second order \textsc{Casimir}-operator.
The bound states transform according to particular
irreducible $SO(d+1)$-representations. The allowed
representations, the explicit form of the bound states 
and their energies have been determined.

We have not discussed the scattering problem.
It is well-known how to extend supersymmetric 
methods from bound to scattering states in
supersymmetric quantum mechanical systems \cite{cooperwipf}.
Thus one may expect that the construction generalizes
to the scattering problem, for which the
non-compact dynamical symmetry group will be $SO(d,1)$.

\textsc{Itzykson} and \textsc{Bander} \cite{bander} distinguished
between the infinitesimal and the global
method to solve the \textsc{Coulomb} problem.
The former is based on the \textsc{Laplace-Runge-Lenz}
vector and is the method used here.
In the second method one performs a stereographic
projection of the $d$-dimensional momentum
space to the unit sphere in $d+1$ dimensions
which in turn implies a $SO(d\!+\!1)$
symmetry group. It would be interesting to
perform a similar global construction for
the supersymmetrized systems.

Every multiplet of the dynamical symmetry group appears 
several times \cite{wipfkirch2} and there is a 
new 'accidental' degeneracy: in higher
dimensions some eigenvalues of
the Hamiltonian appear in many different particle-number sectors. 
It may very well be, that the algebraic
structures discussed in the present 
work have a more natural setting in the language 
of superalgebras or the $SO(d,2)$-setting
in \cite{sudarshan}. We have not
investigated this questions.

There exist earlier results on the supersymmetry 
of both the non-relativistic and relativistic hydrogen 
atom. In \cite{othershatom} the \textsc{Runge-Lenz} 
vector or its relativistic generalization, the 
\textsc{Johnson-Lippmann} operator, enter the
expressions for the supercharges belonging to 
the \emph{ordinary} \textsc{Schr\"odinger-} 
or \textsc{Dirac}-operators with $1/r$ potential.
This should be contrasted with the present work,
where the \textsc{Coulomb}-problem
is only a particular channel of a manifestly supersymmetric
matrix-\textsc{Schr\"odinger} operator. Our \textsc{Hamilton}ians
incorporate both the proton-electron and
the proton-positron  systems as particular
subsectors.

The supercharge (\ref{supercharge})
and super-\textsc{Hamilton}ian (\ref{susyham}) describe
a wide class of supersymmetric systems, ranging
from the supersymmetric oscillator in $d$ dimensions 
to lattice  \textsc{Wess-Zumino}-models with $\cN=1$ or
$\cN=2$ supersymmetries in $2$ dimensions \cite{wipfkirch3}.
In passing we mention, that the supercharge
in $d$ dimension is actually a dimensionally reduced \textsc{Dirac}
operator in $2d$ dimensions. During the reduction process 
the \textsc{Abel}ian gauge potential $A_\mu$ in
$2d$ dimensions transforms into the potential $\chi$
in (\ref{supercharge}), see \cite{wipfkirch3}.

More generally, one may ask for which gauge- and
gravitational background field the \textsc{Dirac}-operator
admits an extended supersymmetry. This question has been
answered in full generality in \cite{kirch1}. 
For example, on a $4$-dimensional hyper-\textsc{K\"ahler} space
with self-dual gauge field the \textsc{Dirac}-operator
admits an $\cN=4$ supersymmetry. The extended supersymmetry
may be used to construct possible zero-modes of
the \textsc{Dirac}-operator. Earlier
results on the supersymmetries of \textsc{Dirac}-type
operators can be found in \cite{holten}, for example.
\textsc{Comtet} and \textsc{Horvathy}
investigated the solutions of the \textsc{Dirac}-equation 
in the hyper-\textsc{K\"ahler} \textsc{Taub-NUT} gravitational 
instanton \cite{horvaty}.
The spin $0$ case can be solved with the help of a \textsc{Kepler}-type 
dynamical symmetry \cite{gibbons} and the fermion case by relating
it to the spin $0$ problem with the help of 
supersymmetry.

\section*{Acknowledgments}
A. Wipf would like to thank the organizers
of the 4th International Symposium 
"Quantum Theory and Symmetries",
and in particular Vladimir Dobrev,
who did a very good job in organizing
this stimulating meeting at the shores of 
the black sea.
We have profited from discussions with and comments from
F. Bruckmann, L. Feher, J.W. van Holten, P. Horvathy,
P. Pisani, I. Sachs and T. Strobl.


\end{document}